\documentclass[acmsmall,screen,nonacm]{acmart}
\setcopyright{none}
\usepackage{nicefrac}
\usepackage{xspace}
\usepackage{acronym}
\usepackage{enumitem}
\usepackage{colortbl}
\usepackage{tabularx}
\usepackage{multirow}
\usepackage{makecell}
\usepackage{adjustbox}
\usepackage{siunitx}
\usepackage{subcaption}
\usepackage{pifont}
\usepackage{booktabs}
\usepackage{graphicx}
\usepackage{listings}
\usepackage[subtle]{savetrees}
\usepackage{microtype}
\usepackage[most]{tcolorbox}
\newcommand{\totalHardwareIfStatements}{\num{3480888}\xspace}
\newcommand{\averageHardwareIfStatementsPerApp}{\num{74}\xspace}
\newcommand{\medianHardwareIfStatementsPerApp}{\num{51}\xspace}
\newcommand{\firstQuartileHardwareIfStatementsPerApp}{\num{16}\xspace}
\newcommand{\thirdQuartileHardwareIfStatementsPerApp}{\num{108}\xspace}
\newcommand{\excludedAppsAfterSootFailure}{\num{3189}\xspace}
\newcommand{\excludedAppsAfterSootFailurePercentage}{\num{6}\%\xspace}
\newcommand{\appsWithHardwareIfStatementsPercentage}{\num{95}\%\xspace}
\newcommand{\hardwareRelatedAPIs}{\num{85}\xspace}
\newcommand{\taxonomyMotivations}{\num{53}\xspace}
\newcommand{\taxonomyDomains}{\num{14}\xspace}

\newcommand{\wcircle}[1]{\ding{\numexpr171 + #1}}
\newcommand{\bcircle}[1]{\ding{\numexpr181 + #1}}

\usepackage[strict]{changepage}
\usepackage{framed}

\definecolor{promptbg}{HTML}{B7E0FF}
\definecolor{promptborder}{HTML}{024CAA}

\definecolor{answerbg}{rgb}{0.85,1,0.85}
\definecolor{answerborder}{rgb}{0.0,0.6,0.30}

\definecolor{examplebg}{RGB}{255,248,220}
\definecolor{exampleborder}{RGB}{230,200,120}

\newenvironment{prompt}{%
  \MakeFramed{\advance\hsize-\width\FrameRestore}%
  \noindent\hspace{-4.55pt}%
  \begin{adjustwidth}{}{7pt}%
}{%
  \end{adjustwidth}%
  \endMakeFramed%
}

\newenvironment{answer}{%
  \MakeFramed{\advance\hsize-\width\FrameRestore}%
  \noindent\hspace{-4.55pt}%
  \begin{adjustwidth}{}{7pt}%
}{%
  \end{adjustwidth}%
  \endMakeFramed%
}

\DeclareCaptionType{codeexample}[Listing][List of Listings]
\newtcblisting{examplelisting}[2]{
  enhanced,
  listing only,
  colback=examplebg,
  colframe=examplebg,
  boxrule=0pt,
  arc=0pt,
  width=\dimexpr\linewidth-2\parindent\relax,
  left skip=\parindent,
  left=5mm,
  right=0.5mm,
  top=0.4mm,
  bottom=0.4mm,
  boxsep=0mm,
  before={\par\smallskip},
  after={\par\captionof{codeexample}{#1}\label{#2}\par\smallskip},
  listing options={
    basicstyle=\ttfamily\footnotesize,
    columns=fullflexible,
    keepspaces=true,
    breaklines=true,
    tabsize=4,
    numbers=left,
    numberstyle=\tiny\color{gray},
    numbersep=6pt,
    stepnumber=1
  }
}

\begin{document}

\title[Hardware-Aware Conditional Code in Android Apps]%
{Hardware-Aware Conditional Code in Android Apps}

\author{Marco Alecci}
\affiliation{\institution{SnT, University of Luxembourg}\country{Luxembourg}}

\author{Jordan Samhi}
\affiliation{\institution{SnT, University of Luxembourg}\country{Luxembourg}}

\author{Li Li}
\affiliation{\institution{School of Software, Beihang University}\country{China}}

\author{Tegawendé F. Bissyandé}
\affiliation{\institution{SnT, University of Luxembourg}\country{Luxembourg}}

\author{Jacques Klein}
\affiliation{\institution{SnT, University of Luxembourg}\country{Luxembourg}}
\renewcommand{\shortauthors}{Alecci et al.}

\begin{abstract}
  Android OS runs on a highly diverse set of devices, spanning entry-level phones and flagships with substantially different hardware characteristics, including memory capacity, processors, displays, cameras, and sensors. Through APIs, apps can query device information at runtime, such as memory, hardware features, and manufacturer, among other device properties, allowing developers to execute different code paths on different devices. Existing studies cover only specific compatibility bugs or brand-specific behavior; however, the broader use of hardware information in app logic has not been comprehensively explored.

We address this question with a large-scale empirical study of hardware-related if statements in Android apps. We statically analyze \num{50000} apps to identify the if statements whose conditions depend on the values of hardware-related APIs, and extract their conditions together with the code of their branches. Our results show that \appsWithHardwareIfStatementsPercentage of successfully analyzed apps contain at least one such statement, with an average of \averageHardwareIfStatementsPerApp per app, which suggests that hardware-related logic is widespread. We then apply an LLM-assisted open-coding process to the extracted code to build a structured taxonomy of the purposes behind hardware-related conditional code. The resulting characterization shows that developers use hardware information for feature availability checks, UI adaptation, emulator detection, device tiering, and resource management, among others.

\end{abstract}

\maketitle

\section{Introduction}
\label{sec:introduction}
Android OS is used across a large and highly diverse ecosystem of devices. Its open-source nature, through the Android Open Source Project (AOSP), allows device manufacturers to adapt and customize the platform for their products~\cite{aospOverview}. As a result, Android applications (apps) execute on a wide variety of devices with different hardware characteristics, e.g., memory capacity, processors, displays, cameras, and sensors, as well as manufacturer-specific system customizations. This heterogeneity is also reflected in the wide range of market segments, from entry-level smartphones priced at around 40~USD in emerging markets to top-tier devices costing around \num{2000}~USD~\cite{dalberg2017accelerating,gsma2021maisha,samsung2025galaxyZFold7}. Already in 2015, more than \num{24000} distinct Android devices from over \num{1200} manufacturers had been observed in the wild~\cite{opensignal2015fragmentation}. More recently, a study published by Google Research reported that Android runs on over 3 billion active devices worldwide and analyzed \num{904} distinct devices from 6 major OEMs~\cite{deng2026android}. Such diversity challenges developers, as the same app may run on devices with substantially different hardware capabilities.

Android provides several mechanisms to help developers support heterogeneous device configurations without explicitly inspecting them in app code. For instance, developers can provide alternative resources for different screen sizes and densities, which Android selects according to the current device configuration~\cite{androidAlternativeResources}. Similarly, Android App Bundles allow Google Play to tailor the APKs delivered to each device~\cite{androidAppBundle}. Rather than delivering the same APK to every device, Google Play can install a base APK together with configuration-specific split APKs containing resources and native libraries for the device's screen density, CPU architecture, and language. 
Nevertheless, developers are not limited to these mechanisms. Indeed, Android exposes APIs that provide information about the device hardware and configuration. For example, \texttt{ActivityManager.isLowRamDevice()} indicates whether the app is running on a low-RAM device~\cite{androidActivityManager}, while APIs such as \texttt{Build.MODEL} expose information about the device model~\cite{androidBuild}. Similar APIs provide access to display properties, available hardware features, supported CPU architectures, camera capabilities, sensors, and other device characteristics. Developers may use this information to make implementation decisions, for example, by disabling memory-intensive functionality on low-RAM devices, selecting implementations based on available hardware features, or applying workarounds to specific device models. Recent work has indeed provided evidence that the same Android app can exhibit different behaviors across different devices~\cite{dong2024device}, but considers only behavior conditioned by device-identity information in Chinese app markets. Our goal is to characterize hardware-related if statements beyond device identity, covering memory, display, capability, camera, sensor, and CPU-related APIs. Other existing studies have investigated Android fragmentation and compatibility issues by focusing on specific cases of device- or API-induced failures and their handling~\cite{wei2016fragmentation,he2018evolution,xia2020compatibility}. However, we still have limited knowledge of how developers explicitly use hardware-related information to make such decisions in their code. In particular, it remains unclear which hardware-related information developers use in the conditions of if statements and, more importantly, what motivates such hardware-related conditional code. This raises our central question: \emph{how do Android developers use hardware-related information to shape the conditions and branches of if statements?}

In this paper, we conduct a large-scale empirical study of hardware-related if statements across \num{50000} Android apps and \hardwareRelatedAPIs Android APIs. We first use static analysis to identify hardware-related APIs whose values influence the conditions of if statements and, for each such statement, extract its condition and the code executed along its two branches.  Our results show that \appsWithHardwareIfStatementsPercentage of successfully analyzed apps contain at least one such statement, with an average of \averageHardwareIfStatementsPerApp statements per app, suggesting that hardware-related logic is widespread in Android apps. We then leverage a Large Language Model (LLM) as an assistant to summarize this code context and support an open-coding process aimed at identifying why developers make decisions based on hardware information. We manually consolidate the resulting reasons into a structured taxonomy of motivations for hardware-related conditional code, including emulator-environment handling, compatibility handling, and resource management, among others. One particularly interesting case is device tiering, whereby apps assign a device to a specific tier or category that can subsequently be reused throughout the codebase to guide later decisions. Overall, our study characterizes how Android developers use hardware information to adapt app behavior across heterogeneous devices, showing that such code can complicate dynamic testing, increase compatibility risks, and lead to different user experiences across devices.

\noindent
\textbf{Contributions.}
We make the following key contributions:
\begin{itemize}[leftmargin=*]
    \item We present the first large-scale empirical characterization of hardware-related if statements in Android apps. By analyzing \num{50000} APKs, we show that \appsWithHardwareIfStatementsPercentage of successfully analyzed apps contain at least one hardware-related if statement, with an average of \averageHardwareIfStatementsPerApp statements per app.
    \item We develop an LLM-assisted open-coding process to identify the motivations behind developers' choices, which are then manually consolidated into a structured taxonomy comprising \taxonomyMotivations motivations organized into \taxonomyDomains domains such as \emph{Emulator and test environment handling}, \emph{Device tiering and memory adaptation}, and \emph{Vendor, model and ROM adaptation}, among others.
    \item We present real-world app examples showing how hardware-related APIs concretely affect app behavior, such as by controlling resource configurations on low-end devices, selecting vendor-specific code paths, and defining custom device tiers that can be reused across the app.
    \item We release our taxonomy and all supporting research artifacts to the community to facilitate reproducibility and further research as described in the \hyperref[sec:dataAvailability]{Data Availability} statement.
\end{itemize}

\section{Background}
\label{sec:background}
In this section, we introduce the main concepts underlying our study.

\noindent
\textbf{Android APKs and App Bundles.}
Android apps are commonly distributed as Android Package (APK) files, which contain the app code, resources, manifest, and native libraries required for execution. In 2018, Google introduced the Android App Bundle (AAB) as a new publishing format for Android apps, and since August 2021, new apps published on Google Play are required to use it~\cite{androidAppBundle}. Unlike APKs, AAB files are not directly installed on devices; rather, developers upload an app bundle to Google Play, which then generates and delivers a set of APKs tailored to the target device configuration. This typically includes a base APK together with configuration-specific split APKs containing resources or native libraries selected according to characteristics such as screen density, CPU architecture, and language~\cite{androidAppBundle}. This mechanism allows part of the device heterogeneity to be handled during app delivery rather than explicitly in app code. In this work, we focus instead on hardware-related differences that are handled through conditional code in the analyzed code of the base APK, while differences introduced exclusively through split APK delivery are outside the scope of this study and are left for future work.

\noindent
\textbf{Hardware-Related Information in Android Apps.}
Android exposes a variety of device-related information through APIs. Although we collectively refer to these APIs as \emph{hardware-related}, they describe not only physical hardware and its capabilities, but also device identity and configuration information, such as the model, manufacturer, or supported CPU architectures. For example, \texttt{FingerprintManager.isHardwareDetected()} indicates whether fingerprint hardware is available on the device~\cite{androidFingerprintManager}, while \texttt{ActivityManager.MemoryInfo.totalMem} reports the total amount of memory available on the device~\cite{androidActivityManager}.
For the remainder of this paper, we define an Android API as \emph{hardware-related} when it exposes device-level information about hardware, capabilities, capacity, identity, compatibility, or configuration that developers can use to differentiate code behavior across devices, rather than merely reporting a transient operational state. As a clarifying example, APIs that report the remaining battery percentage (despite being related to a hardware property) do not fit this definition because they describe the device's current charge state rather than the characteristics considered within the scope of this study.

\section{Experimental Setup}
\label{sec:experimentalSetup}
This section presents the dataset, the list of hardware-related APIs, our two-phase methodology, and the implementation details.
\subsection{Dataset}
\label{sec:dataset}
Our dataset consists of \num{50000} recent Android apps randomly selected from AndroZoo~\cite{allix2016androzoo,alecci2024androzoo}, considering apps newly collected from Google Play between August 2024 and August 2026. This dataset provides a realistic and up-to-date view of current development practices, as it includes production apps that were available on Google Play during this period. To the best of our knowledge, this is the largest study to date on hardware-related code in Android apps in terms of the number of analyzed apps.

\subsection{Hardware-Related APIs}
\label{sec:hardwareRelatedAPIs}
The Android API surface contains more than \num{45000} documented APIs~\cite{alecci2026taskflow,samhi2023codoc}, making an exhaustive manual inspection extremely challenging. We therefore adopted a seed-and-expand procedure to construct a focused manually validated list of hardware-related APIs. We first asked ChatGPT~\cite{openai2023chatgpt} to suggest Android APIs that may expose device hardware, capabilities, capacity, or identity information, using the definition introduced in Section~\ref{sec:background}, and then manually inspected the official Android documentation for every suggested API, retaining it only when it satisfied our definition. To further expand the list, we also examined semantically related APIs within the same Android classes and packages. For example, after \texttt{android.os.Build.MODEL} was suggested, we systematically inspected the other APIs exposed by \texttt{android.os.Build}. The entire process resulted in a final list of \hardwareRelatedAPIs hardware-related APIs organized into \num{11} topics, such as \emph{Storage Capacity} and \emph{Build Identity}. For instance, it includes \texttt{PackageManager.hasSystemFeature()} for capability checks, \texttt{Build.MODEL} and \texttt{SUPPORTED\_ABIS} for device identity and native compatibility, camera-characteristic keys, and \texttt{ActivityManager.isLowRamDevice()} for memory-aware behavior. Table~\ref{tab:hwApiList} provides an overview of the \num{11} topics and the number of APIs associated with each of them, while the complete list is available in our repository. It is important to note that our list is not intended to be a complete enumeration of every hardware-related Android API; rather, it represents a trade-off that allows us to characterize the phenomenon at the ecosystem scale while accepting that some more complex hardware-dependent behaviors may remain outside our analysis. Our study intentionally prioritizes scalability over exhaustive program understanding, as further discussed in Section~\ref{sec:limitations}.

\begin{table}[ht]
  \caption{Hardware-related Android APIs summary. The full detailed list can be found in our repository.}
  \label{tab:hwApiList}
  \centering
  \small
  \begin{adjustbox}{width=0.75\linewidth}
    \begin{tabular}{@{}lr@{\hspace{3em}}lr@{}}
      \toprule
      Topic & N. APIs &
      Topic & N. APIs \\
      \midrule
      Biometric hardware & 5 &
      Graphics and GPU capability & 3 \\

      Build identity & 17 &
      Memory capacity & 4 \\

      Camera characteristics & 5 &
      Package-manager hardware features & 26 \\

      CPU, ABI, and SoC & 6 &
      Sensors & 3 \\

      Display and screen characteristics & 8 &
      Storage capacity & 4 \\

      Telephony hardware & 4 &
      & \\

      \midrule
      & & 
      \textbf{Total} & \textbf{\hardwareRelatedAPIs} \\
      \bottomrule
    \end{tabular}
  \end{adjustbox}
\end{table}

\subsection{Methodology}
\label{sec:methodology}
\begin{figure*}[t]
  \centering
  \includegraphics[width=0.95\linewidth]{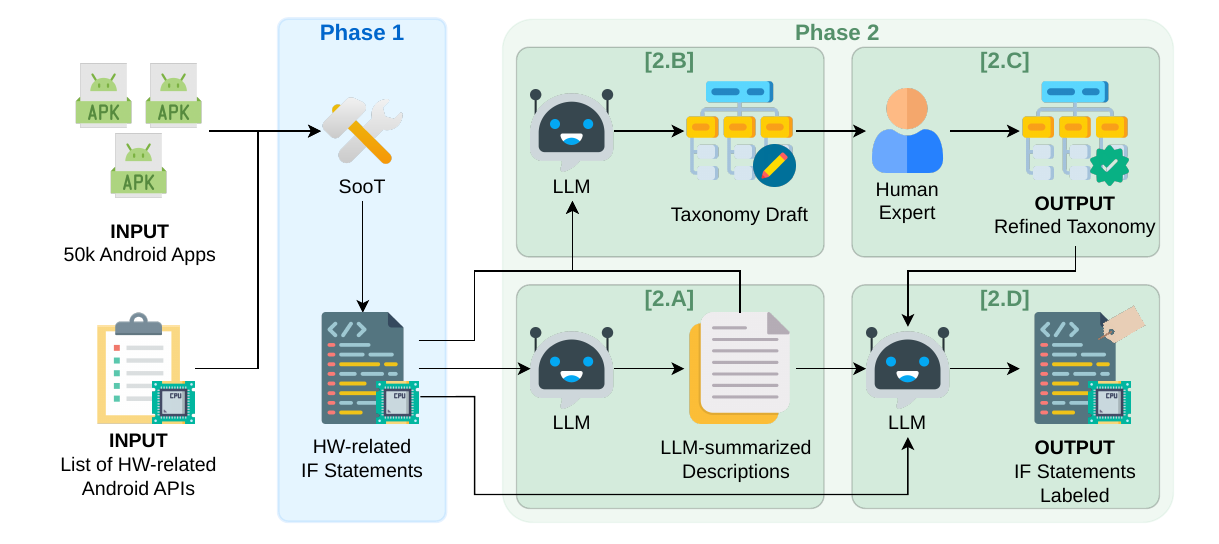}
  \caption{Approach Overview.}
  \label{fig:approachOverview}
\end{figure*}

Our goal is to build a taxonomy that characterizes why Android apps use hardware-related information in if statements. To this end, we designed a two-phase pipeline consisting of \bcircle{1} Extraction of Hardware-Related If Statements and \bcircle{2} LLM-assisted Open Coding, which are described in detail below. An overview of the experimental pipeline is shown in Figure~\ref{fig:approachOverview}.

\subsubsection{Phase 1: Extraction of Hardware-Related If Statements}
Since the source code of Android apps is generally not available, static analyses commonly operate directly on APK bytecode. In the first phase of our experimental pipeline, we rely on Soot~\cite{vallee2010soot} to load each APK and represent its bytecode using Jimple, i.e., Soot's typed three-address intermediate representation. One advantage of Jimple is that it relies on a small, regular set of 15 statement types, such as \texttt{AssignStmt}, \texttt{GotoStmt}, and \texttt{IfStmt}, among others, thereby allowing bytecode instructions to be expressed in a uniform representation.
We then filter out all code belonging to third-party libraries relying on AndroLibZoo~\cite{samhi2024androlibzoo} in order to focus on non-library code and avoid counting identical if statements across thousands of APKs. We then inspect the Jimple representation of the remaining code to identify all \texttt{IfStmt} instructions and determine, for each of them, whether its condition depends on one of the hardware-related APIs in our list. Specifically, we retain if statements where either a hardware-related API appears directly in the \texttt{if} condition, or a local variable used in the condition is assigned from hardware-related information within the same method. Whenever a match is found, we collect \wcircle{1} the hardware-related API signature, \wcircle{2} the enclosing class and method identifiers, \wcircle{3} the \texttt{if} condition, and \wcircle{4} the code associated with its \texttt{if} and \texttt{else} branches. 
One important point is that our goal is not to perform the most complete whole-program analysis possible, but rather to identify a large number of hardware-related if statements across a large set of real-world apps. We therefore deliberately keep the analysis lightweight and intra-procedural: when a branch invokes another method, we do not recursively analyze that method to determine its eventual effects, nor do we track whether method parameters contain hardware-related information propagated from other methods. This design favors scalability over completeness and enables the large-scale analysis performed in our study, while its implications are further discussed in Section~\ref{sec:limitations}. \excludedAppsAfterSootFailure apps (\excludedAppsAfterSootFailurePercentage) could not be successfully processed by Soot (e.g., due to type-resolution failures, unresolved references, malformed bytecode, and similar errors) and were excluded from the subsequent analysis. The outcome of this first phase is a set of \totalHardwareIfStatements hardware-related if statements, averaging \averageHardwareIfStatementsPerApp per successfully analyzed app, which constitute the input to Phase~2.

\subsubsection{Phase 2: LLM-Assisted Open Coding.}
Once the hardware-related if statements have been identified, the next step is to understand why developers use such information to control app behavior. Inspired by prior work on thematic synthesis, taxonomy development, and qualitative analysis in software engineering, as well as recent work that relies on LLMs to support open coding and inductive qualitative analysis, we designed an LLM-assisted open-coding process comprising four subphases~\cite{cruzes2011thematic,stol2016grounded,usman2017taxonomies,mathis2024inductive,lecca2025applications}. It is important to note that LLMs are used only as a scalable aid to summarize the supplied code evidence and support the open-coding process and final assignment, since a fully manual, record-by-record qualitative analysis of all if statements would be extremely challenging given the scale of the dataset.
Before describing the four subphases, it is important to clarify how we make the semantic analysis feasible at the scale of our dataset. Phase 1 identifies \totalHardwareIfStatements hardware-related if statements across our dataset, making an exhaustive semantic analysis extremely challenging and impractical, even with the support of an LLM, in terms of both computational resources and execution time. Rather than analyzing the entire population, we therefore adopt a per-signature random sampling strategy in which we independently draw, for each hardware-related API signature, a sample using a 99\% confidence level and a 5\% margin of error. Similar sampling-based strategies are commonly used in empirical software engineering when exhaustive inspection of large datasets is impractical~\cite{li2026automatically,guglielmi2024help,alecci2025damflow,zhang2025killing,wang2024makes,alecci2026taskflow,alecci2024improving,alecci2026geotwins,ferrari2026evaluating, alecci2025evaluating}. This per-signature sampling procedure prevents highly frequent APIs from dominating the analysis and yields a total of \num{32732} hardware-related if statements, which are then used for our taxonomy construction process. We now describe the four subphases of our open-coding process.

\noindent
\bcircle{1} \textbf{Phase 2-a: LLM-Summarized Descriptions.}
For each if statement, we use the LLM to compare the behavior of the two branches and summarize their behavior into an evidence-bound description using the following prompt.

\begin{prompt}
\small
\textbf{Prompt: } \emph{You are a senior Android reverse engineer. Analyze the following Jimple IF statement from an Android app. Explain what the specified hardware-related API is used for in this code and how the IF/else branches use its result. Base your answer only on the supplied code context. If the context is insufficient, say so explicitly. Do not invent app behavior. Return a concise plain-text explanation (maximum three sentences).} 
\end{prompt}

\noindent
Together with each Jimple if statement, we provide the LLM with the enclosing class and method identifiers, the hardware-related API signature, as well as the code fragments associated with both branches. The following is an LLM-summarized description for a check of \texttt{ActivityManager.isLow\-RamDevice()}, which is used to select the default fraction of available memory allocated to an in-memory cache:

\begin{prompt}
\small
\textbf{LLM-summarized description.} \emph{The method \texttt{isLowRamDevice()} is used to detect whether the device has limited memory. If the device is not a low-RAM device, the function returns \texttt{0.20}; if it is a low-RAM device, it returns \texttt{0.15}, meaning a smaller default cache percentage is used. Thus, the IF/ELSE branches adjust the cache size based on the device's memory capacity.}
\end{prompt}

\noindent
An LLM-summarized description is obtained in this way for each sampled if statement.

\noindent
\bcircle{2} \textbf{Phase 2-b: Taxonomy Exploratory.}
Once an LLM-summarized description is available for each if statement, the next step is to build a taxonomy of \emph{motivations}, i.e., the reasons why developers make decisions based on hardware-related information (e.g., \emph{emulator detection} or \emph{storage capacity scaling}). The \num{32732} if statements, together with their corresponding LLM-summarized descriptions, are processed in batches of 50. For each batch, the LLM supports the exploratory identification of potential motivations and is free to suggest new concepts. This flexibility is useful during exploration, but it can also lead the model to suggest different labels for substantially the same underlying behavior across separate batches. The exploratory pass yields \num{1423} candidate motivation proposals, which therefore do not yet constitute the final taxonomy. For instance, \emph{emulator detection via build properties} and \emph{emulator detection} were proposed separately despite describing substantially the same behavior, highlighting the need for a subsequent manual consolidation step to merge overlapping concepts and resolve naming inconsistencies.

\noindent
\bcircle{3} \textbf{Phase 2-c: Manual Taxonomy Consolidation.}
One of the authors manually reviewed the candidate motivations suggested by the LLM during Phase~2-b to resolve overlapping concepts and naming inconsistencies, while also grouping related motivations into higher-level domains. This human pass produces a fixed taxonomy comprising \taxonomyMotivations motivations organized into \taxonomyDomains high-level domains. For example, the domain \emph{Emulator and test environment handling} encompasses both motivations for detecting emulator environments and motivations for applying emulator- or test-specific behavior. The complete taxonomy is available in our repository and is presented later in Section~\ref{sec:experimentalResults}.

\noindent
\bcircle{4} \textbf{Phase 2-d: Taxonomy Final Assignment.}
In the final pass, the LLM is asked to assign exactly one motivation from the fixed taxonomy to each if statement, without the possibility of creating, renaming, or combining categories. This prevents the taxonomy from changing during classification and ensures that each final label remains linked to the corresponding code evidence. To mitigate the LLM's non-determinism, we repeat the classification query 5 times for each if statement and retain the motivation selected in the majority of the runs. Finally, one of the authors manually checked the labels assigned to a random sample of \num{380} final assignments, sized for a 95\% confidence level and a 5\% margin of error, against the underlying Jimple evidence and LLM-summarized descriptions; this check identified no labeling errors.

The output of Phase~2 is therefore a fixed taxonomy comprising \taxonomyMotivations motivations organized into \taxonomyDomains high-level domains, together with a complete assignment of the \num{32732} sampled if statements to one specific motivation and its corresponding domain. Potential threats to the validity of the taxonomy and of the overall coding process, including the use of LLM-summarized descriptions, are discussed in Section~\ref{sec:limitations}.

\subsection{Implementation Details}
\label{sec:implementationDetails}
For Phase~1, we implemented a custom Java analysis based on Soot~\cite{vallee2010soot} to process the APKs and extract hardware-related if statements. The analysis was executed using eight parallel workers on a machine equipped with two Intel Xeon Silver 4416+ processors, 251~GiB of RAM, and Ubuntu~22.04.5~LTS.
For the LLM-assisted analysis, we rely on the open-source \texttt{gpt-oss:20b} model~\cite{openai2025gptoss}, executed locally on a NVIDIA RTX 5000 Ada Generation GPU with 32 GiB of VRAM. Running the model locally allows us to keep the computational and financial costs of the experiments manageable while enabling other researchers to rerun the analysis with the same open model, prompts, sampling seed, and configuration. The prompt used for the description phase is reported above, while the remaining prompts, model parameters, and LLM configuration are available in our repository. The choice of the LLM and its implications are further discussed in Section~\ref{sec:limitations}.
Overall, the complete experimental pipeline required 15 days, 21 hours, and 43 minutes on our hardware. Table~\ref{tab:approachTimeElapsed} reports a detailed breakdown of the runtime of each phase and subphase. 

\begin{table}[ht]
    \caption{Detailed breakdown of the runtime for each phase of the experimental pipeline.}
  \label{tab:approachTimeElapsed}
  \centering
  \small
  \begin{adjustbox}{width=0.6\linewidth}
    \begin{tabular}{@{}llll@{}}
      \toprule
      \multicolumn{1}{c}{\textbf{Phase}} & \multicolumn{1}{c}{\textbf{Subphase}} & \multicolumn{2}{c}{\textbf{Runtime}} \\
      \midrule
      Phase 1 
        & Hardware-related API extraction 
        & \multicolumn{2}{c}{12 days 21 h} \\
      \midrule
      \multirow{3}{*}{Phase 2}
        & LLM-summarized description 
        & 27 h 18 min 
        & \multirow{3}{*}{3 days 43 min} \\
        & Taxonomy discovery 
        & 7 h 53 min 
        & \\
        & Frozen-taxonomy assignment 
        & 37 h 32 min 
        & \\
      \midrule
      \textbf{Total} 
        & -- 
        & \multicolumn{2}{c}{\textbf{15 days 21 h 43 min}} \\
      \bottomrule
    \end{tabular}
  \end{adjustbox}
\end{table}

Phase~1 accounts for most of the execution time, requiring 12 days and 21 hours to analyze the \num{50000} APKs. The LLM stages required approximately 72 hours and 43 minutes to process the sampled \num{32732} if statements. Since the complete population contains \totalHardwareIfStatements if statements, approximately 106 times as many as the analyzed sample, processing all if statements with the same configuration would have required substantially more computational time, further justifying our sampling strategy.

\section{Experimental Results}
\label{sec:experimentalResults}

In this section, we present the results of our empirical evaluation. We organize the analysis around the following Research Questions (RQs):

\begin{itemize}[leftmargin=*]
    \item \textbf{RQ1:} How prevalent are hardware-related if statements in Android apps, and which hardware-related APIs are most frequently used?
    \item \textbf{RQ2:} What are the main motivations behind hardware-related if statements, and how do they shape app behavior across different devices?
    \item \textbf{RQ3:} How do hardware-related APIs map to the motivations identified in our taxonomy?
\end{itemize}

\subsection{RQ1: Usage of Hardware-Related APIs}
\label{sec:rq1}
In this RQ, we aim to provide an overview of the prevalence of hardware-related if statements and the distribution of the APIs used in them by analyzing the output of Phase~1 (see Section~\ref{sec:methodology}), i.e., all the \totalHardwareIfStatements hardware-related if statements identified across the entire dataset. 

\begin{figure}[ht]
  \centering
  \includegraphics[width=0.6\linewidth]{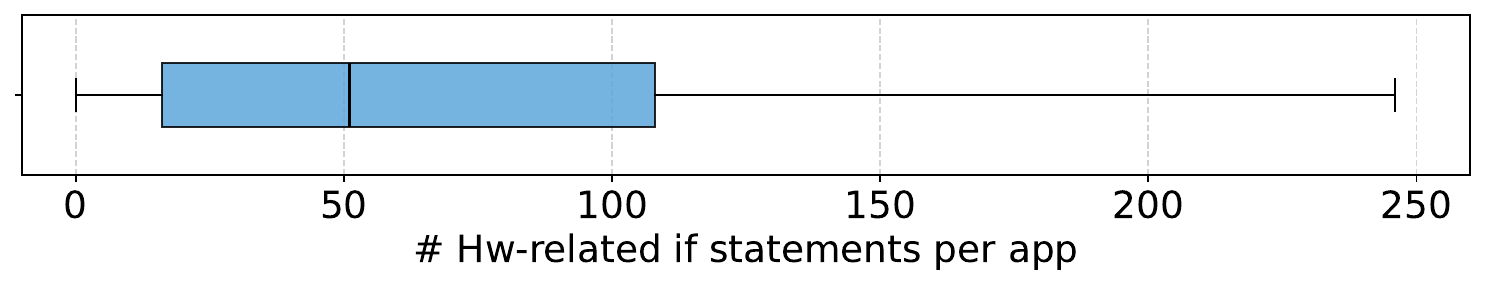}
  \caption{Distribution of hardware-related if statements per app.}
  \label{fig:rq1OverallUsageDistribution}
\end{figure}

Overall, \appsWithHardwareIfStatementsPercentage of the successfully analyzed apps contain at least one hardware-related if statement, with an average of \averageHardwareIfStatementsPerApp statements per app. This prevalence indicates that hardware-related if statements are not confined to a small number of specialized apps or isolated compatibility patches, but are instead a recurring implementation mechanism across Android apps. Figure~\ref{fig:rq1OverallUsageDistribution} further characterizes this phenomenon by showing the distribution of hardware-related if statements per app.

\noindent
The median app contains \medianHardwareIfStatementsPerApp hardware-related if statements, while the interquartile range spans from \firstQuartileHardwareIfStatementsPerApp to \thirdQuartileHardwareIfStatementsPerApp. The right-skewed distribution shows that, although this logic is common across apps, a smaller subset contains substantially more hardware-related if statements than the typical app. We now focus on investigating which individual hardware-related APIs are most widely adopted across successfully analyzed apps. Table~\ref{tab:rq1Top20AppAdoption} reports the 20 most widely adopted hardware-related APIs, ranked by the number of successfully analyzed apps in which each API occurs at least once in an if statement.

\begin{table*}[ht]
  \centering
  \caption{The 20 hardware-related APIs used in if statements by the largest number of apps.}
  \label{tab:rq1Top20AppAdoption}
  \scriptsize
  \begin{adjustbox}{width=0.9\linewidth}
  \begin{tabular}{ll@{\hspace{1.8cm}}ll}
    \toprule
    \textbf{Hardware reference} & \textbf{Apps} &
    \textbf{Hardware reference} & \textbf{Apps} \\
    \midrule
    \texttt{Build.MANUFACTURER} & \num{31617} [\num{67.5}\%] &
    \texttt{Build.HARDWARE} & \num{20856} [\num{44.6}\%] \\
    \texttt{Resources.getDisplayMetrics()} & \num{31433} [\num{67.1}\%] &
    \texttt{Configuration.smallestScreenWidthDp} & \num{20417} [\num{43.6}\%] \\
    \texttt{PackageManager.hasSystemFeature()} & \num{30449} [\num{65.0}\%] &
    \texttt{DisplayMetrics.heightPixels} & \num{20153} [\num{43.1}\%] \\
    \texttt{Build.MODEL} & \num{30426} [\num{65.0}\%] &
    \texttt{SensorManager.getDefaultSensor()} & \num{19599} [\num{41.9}\%] \\
    \texttt{Build.TYPE} & \num{27148} [\num{58.0}\%] &
    \texttt{DisplayMetrics.densityDpi} & \num{18092} [\num{38.6}\%] \\
    \texttt{Build.FINGERPRINT} & \num{24115} [\num{51.5}\%] &
    \texttt{Build.TAGS} & \num{16092} [\num{34.4}\%] \\
    \texttt{Build.DEVICE} & \num{23884} [\num{51.0}\%] &
    \texttt{Build.SUPPORTED\_ABIS} & \num{14941} [\num{31.9}\%] \\
    \texttt{Build.BRAND} & \num{22328} [\num{47.7}\%] &
    \texttt{Configuration.screenLayout} & \num{14727} [\num{31.5}\%] \\
    \texttt{DisplayMetrics.widthPixels} & \num{22019} [\num{47.0}\%] &
    \texttt{ActivityManager.isLowRamDevice()} & \num{12600} [\num{26.9}\%] \\
    \texttt{Build.PRODUCT} & \num{21274} [\num{45.4}\%] &
    \texttt{CameraCharacteristics.get(Key)} & \num{12545} [\num{26.8}\%] \\
    \bottomrule
  \end{tabular}
  \end{adjustbox}
\end{table*}

\begin{figure*}[ht]
  \centering
  \includegraphics[width=0.85\linewidth]{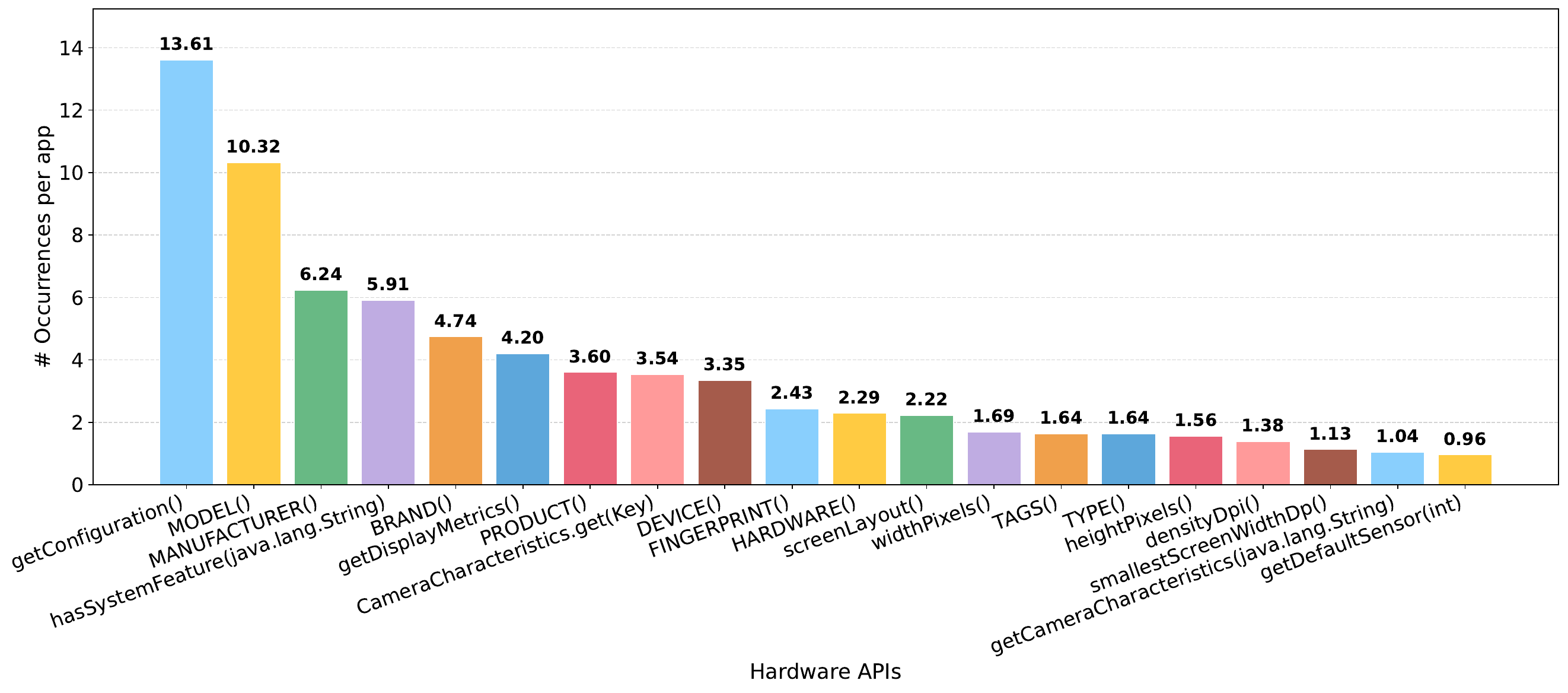}
  \caption{Top 20 hardware-related references ranked by their average number of occurrences per app.}
  \label{fig:top20AverageUsage}
\end{figure*}

From Table~\ref{tab:rq1Top20AppAdoption}, it can be observed that device-identity information is among the most widely used hardware-related information in if statements. Interestingly, \texttt{Build.MANUFACTURER} is the most widely adopted reference, appearing in \num{67.5}\% of successfully analyzed apps. Display-related information is also common, with several APIs appearing in a large fraction of successfully analyzed apps. Finally, even more specialized APIs, such as \texttt{ActivityManager.isLowRamDevice()}, are used in a substantial number of apps, appearing in \num{26.9}\% of successfully analyzed apps.
Figure~\ref{fig:top20AverageUsage} shifts the focus from cross-app adoption to within-app reuse by ranking hardware-related APIs according to their average number of occurrences per app. 

\begin{figure*}[ht]
  \centering
  \includegraphics[width=0.85\linewidth]{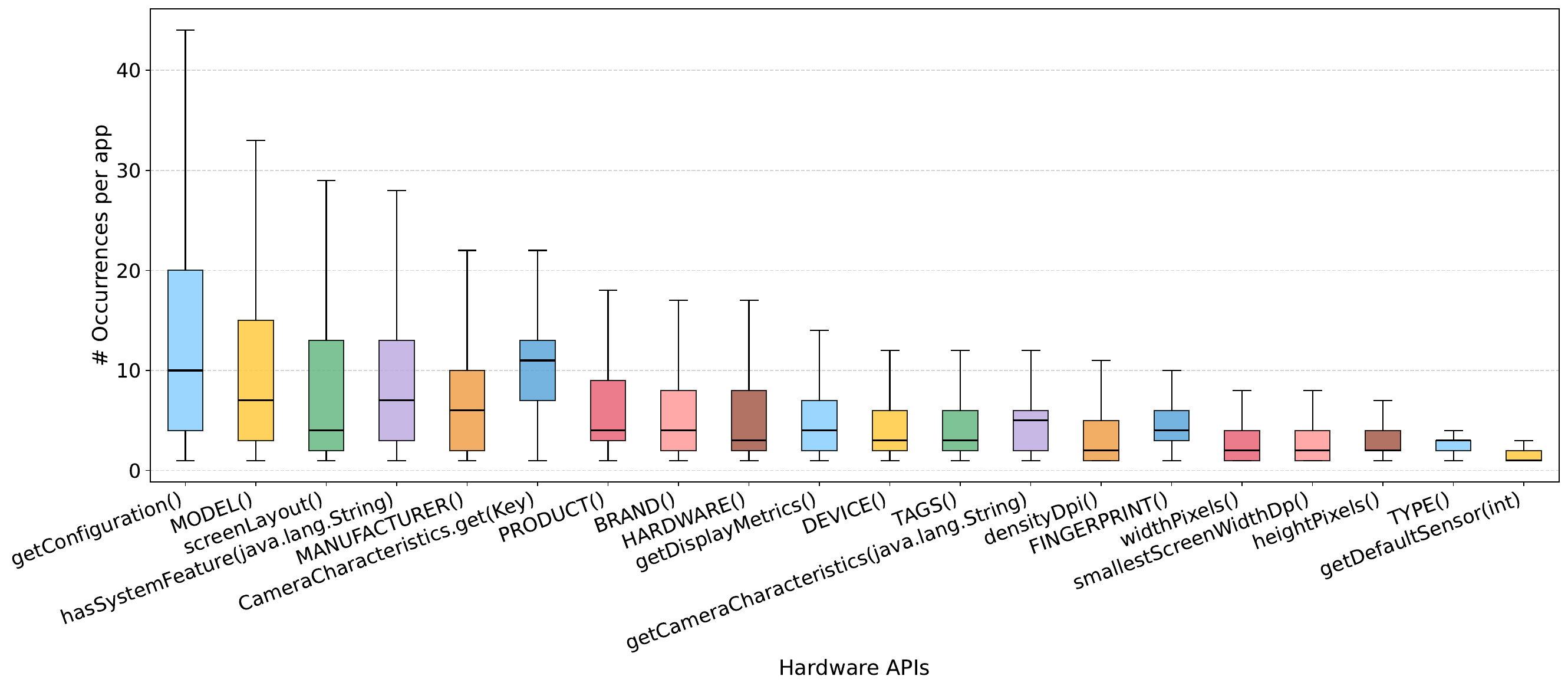}
  \caption{Distribution of occurrences per app for the 20 most frequently used hardware-related references.}
  \label{fig:top20UsageDistribution}
\end{figure*}

Interestingly, \texttt{Build.MODEL} is the most frequently reused reference, with an average of 10.32 occurrences per app. Device-identity information remains prominent in this view as well. Compared with Table~\ref{tab:rq1Top20AppAdoption}, this shows that some APIs are not only widely adopted across apps, but are also repeatedly used in multiple hardware-related if statements within the same app. Camera-related references also appear among the most reused APIs, although their overall adoption across apps is more limited, suggesting a more specialized usage pattern.
Finally, Figure~\ref{fig:top20UsageDistribution} shows the distributions of the number of occurrences per app for the most frequently reused hardware-related references. 

\texttt{Build.MODEL} exhibits a particularly broad distribution, with a median of 7 occurrences among apps that use it, compared with an overall average of 10.32 occurrences per app. This difference suggests that a subset of apps reuses \texttt{Build.MODEL} particularly intensively. A similar pattern can be observed for several other references, where the median among apps that actually use the API differs noticeably from the overall average, highlighting substantial variation in within-app reuse.

Taken together, the table and figures show that hardware-related if statements are widespread across Android apps, rather than being confined to a small set of apps or isolated compatibility patches. Within this widespread use, some APIs are particularly prevalent: device-identity information, such as the device model and manufacturer, and display-related references dominate the results. In contrast, several camera-related references occur more frequently within the apps that use them, but are present in a comparatively smaller share of Android apps. The implications of these usage patterns are further discussed in Section~\ref{sec:discussion}.

\begin{answer}
\textbf{Answer to RQ1.}
Hardware-related if statements are widespread across Android apps: \appsWithHardwareIfStatementsPercentage of the apps contain at least one, with an average of \averageHardwareIfStatementsPerApp statements per app. Device-identity and display APIs are particularly prevalent, while more specialized APIs, such as memory-related or camera-related APIs, are used by fewer apps.
\end{answer}

\subsection{RQ2: Motivations Behind Hardware-Related If Statements}
\label{sec:rq2}
\begin{figure*}[ht]
  \centering
  \includegraphics[width=0.9\linewidth]{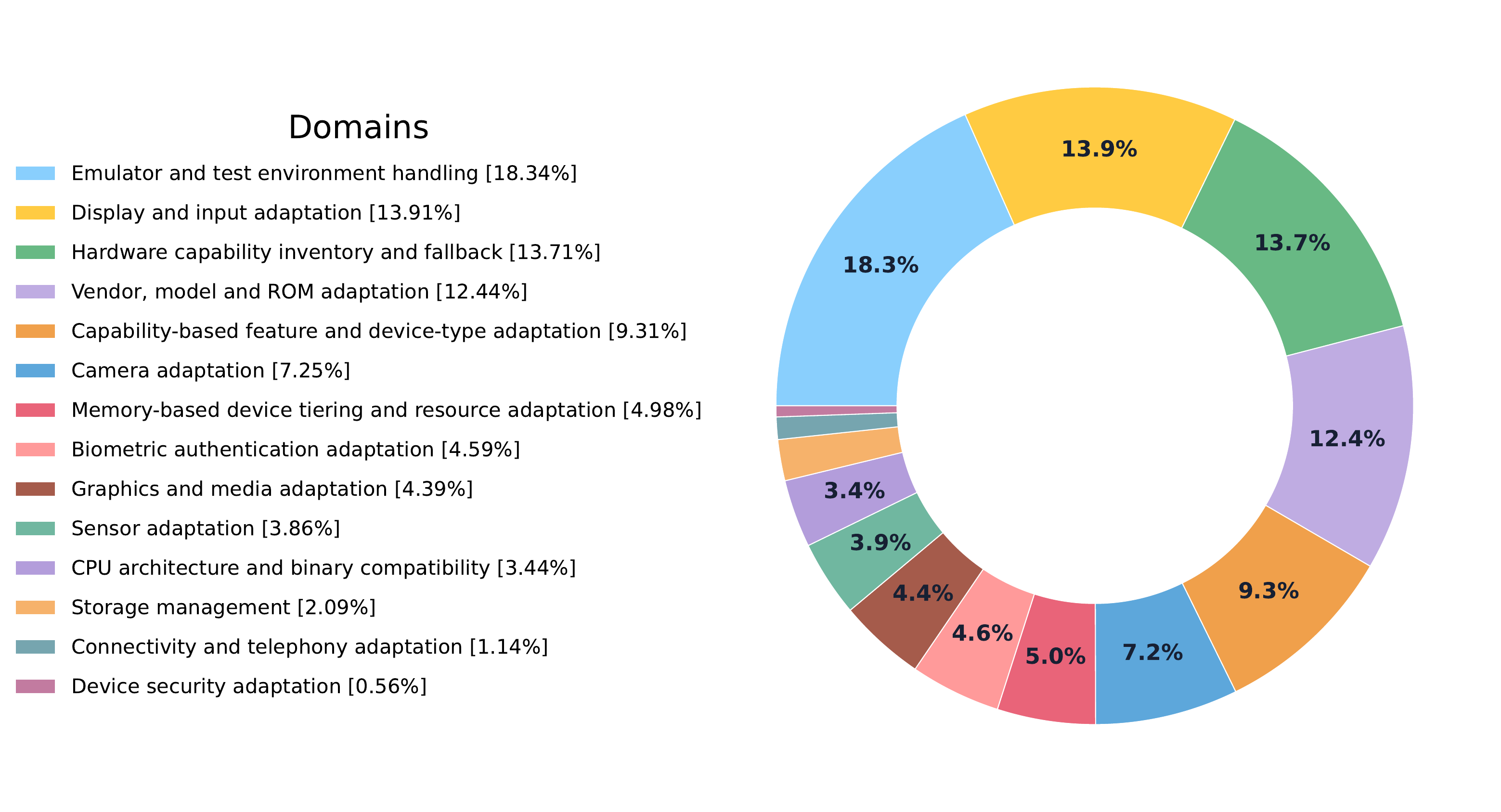}
  \caption{Distribution of sampled hardware-related if statements across taxonomy domains.}
  \label{fig:rq2DomainDistribution}
\end{figure*}

We next investigate the motivations behind hardware-related if statements. As described in Section~\ref{sec:methodology}, our LLM-assisted open-coding process produces a fixed taxonomy comprising \taxonomyMotivations motivations organized into \taxonomyDomains high-level domains. In this RQ, we analyze the distribution of these motivations and highlight representative use cases that illustrate how hardware information influences app behavior. Figure~\ref{fig:rq2DomainDistribution} summarizes the distribution of hardware-related if statements across the taxonomy domains.

The distribution is broad: the largest domain, \emph{Emulator and test environment handling}, accounts for \num{18.3}\% of the sample, followed by \emph{Display and input adaptation} (\num{13.9}\%), \emph{Hardware capability inventory and fallback} (\num{13.7}\%), and \emph{Vendor, model and ROM adaptation} (\num{12.4}\%). There is no single dominant domain, with the remaining domains accounting for progressively smaller shares of the sample. This distribution suggests that hardware-related if statements are used to support a wide range of adaptation strategies, rather than being concentrated in a small number of use cases. Each domain groups closely related motivations that capture different aspects of the same underlying adaptation strategy. For example, \emph{Emulator and test environment handling} includes both detecting that an app is running on an emulator and activating behavior specifically intended for emulator or test environments. In order to observe this distribution at a finer granularity, Table~\ref{tab:rq2Top20Motivations} reports the \num{20} most frequent individual motivations for hardware-related if statements. The complete distribution is available in our repository.

\begin{table*}[ht]
  \centering
  \caption{Top 20 motivations ranked by their share of hardware-related if statements.}
  \label{tab:rq2Top20Motivations}
  \footnotesize
  \begin{adjustbox}{width=0.85\linewidth}
  \begin{tabular}{lr}
    \toprule
    \textbf{Motivation [Domain]} & \textbf{Occurrences [\%]} \\
    \midrule
    Detect emulator environment [Emulator and test environment handling] & \num{13.1}\% \\
    Select layout by screen size and form factor [Display and input adaptation] & \num{6.9}\% \\
    Build hardware capability inventory [Hardware capability inventory and fallback] & \num{6.7}\% \\
    Enable/disable functionality by hardware capability [Capability-based feature and device-type adaptation] & \num{6.6}\% \\
    Select available camera and implementation [Camera adaptation] & \num{5.0}\% \\
    Apply test- and development-specific behavior [Emulator and test environment handling] & \num{4.5}\% \\
    Enable biometric authentication when ready [Biometric authentication adaptation] & \num{4.3}\% \\
    Expose normalized hardware information [Hardware capability inventory and fallback] & \num{4.0}\% \\
    Adapt interface scale and resources to display density [Display and input adaptation] & \num{3.8}\% \\
    Identify vendor or device model [Vendor, model and ROM adaptation] & \num{3.8}\% \\
    Apply vendor- or ROM-specific behavior [Vendor, model and ROM adaptation] & \num{3.1}\% \\
    Apply device-specific workaround [Vendor, model and ROM adaptation] & \num{2.9}\% \\
    Activate sensor-dependent functionality [Sensor adaptation] & \num{2.9}\% \\
    Detect custom ROM environment [Vendor, model and ROM adaptation] & \num{2.6}\% \\
    Return safe result for unavailable hardware information [Hardware capability inventory and fallback] & \num{2.4}\% \\
    Select architecture-compatible binary or interface [CPU architecture and binary compatibility] & \num{2.4}\% \\
    Gate graphics functionality by API capability [Graphics and media adaptation] & \num{2.3}\% \\
    Configure interface for display orientation [Display and input adaptation] & \num{2.3}\% \\
    Resize buffers and allocation limits [Memory-based device tiering and resource adaptation] & \num{2.3}\% \\
    Select supported camera stream configuration [Camera adaptation] & \num{1.6}\% \\
    \bottomrule
  \end{tabular}
  \end{adjustbox}
\end{table*}

As observed for the domains, the distribution is not dominated by a single motivation. The three most frequent motivations are \emph{Detect emulator environment} (\num{13.1}\%), \emph{Select layout by screen size and form factor} (\num{6.9}\%), and \emph{Build hardware capability inventory} (\num{6.7}\%). Although emulator detection is the most common individual motivation, it accounts for only a modest share of the sample, reflecting the diversity of hardware-related adaptation decisions. It is also important to note that although some motivations may initially appear similar, they actually capture slightly distinct decisions. For instance, \emph{Apply vendor- or ROM-specific behavior} describes behavior selected for a vendor ecosystem or a customized Android distribution, whereas \emph{Apply device-specific workaround} targets a known issue of one or more particular device models.
The identified motivations also correspond to concrete implementation patterns in the analyzed apps. We observed checks of \texttt{Build.MODEL} against lists of known devices to apply camera-related workarounds; checks of \texttt{StatFs.getAvailableBytes()} to avoid or defer storage-consuming operations; checks of \texttt{Build.MANUFACTURER} or \texttt{Build.BRAND} to select vendor-specific features or identity mechanisms; checks of \texttt{PackageManager.hasSystemFeature()} to enable or disable a hardware-dependent feature; and checks of \texttt{ActivityManager.isLowRamDevice()} or available memory to constrain caches and graphics-related allocations.
Together, these cases show that the taxonomy captures both broad adaptation strategies (e.g., adapting code to low-end devices or detecting emulator environments) and targeted compatibility decisions (e.g., device-specific workarounds). Detailed code-level examples from the real-world apps we analyzed are presented and discussed later in Section~\ref{sec:discussion}.
One particularly interesting scenario is \emph{memory-based device tiering}, in which developers assign a specific level or tier to a device, suggesting that this classification may be reused elsewhere in the app. For example, one app compares \texttt{ActivityManager.MemoryInfo.totalMem} against \SI{2}{\giga\byte}: devices at or below the threshold retain their existing capability code, whereas devices above it receive an incremented code. This motivates future work on how developers encode and propagate device-tier policies, and whether such classifications remain meaningful as Android hardware continues to evolve.

\begin{answer}
\textbf{Answer to RQ2.}
Hardware-related if statements support diverse adaptation strategies, with no single domain or motivation dominating the results. The identified motivations range from broad adaptations, such as resource management on low-end devices, to targeted compatibility workarounds; \emph{memory-based device tiering} further suggests that hardware-derived information can be translated into reusable app-level policies.
\end{answer}

\subsection{RQ3: Mapping APIs to Motivations}
\label{sec:rq3}
\begin{figure*}[ht]
  \centering
  \includegraphics[width=0.95\linewidth]{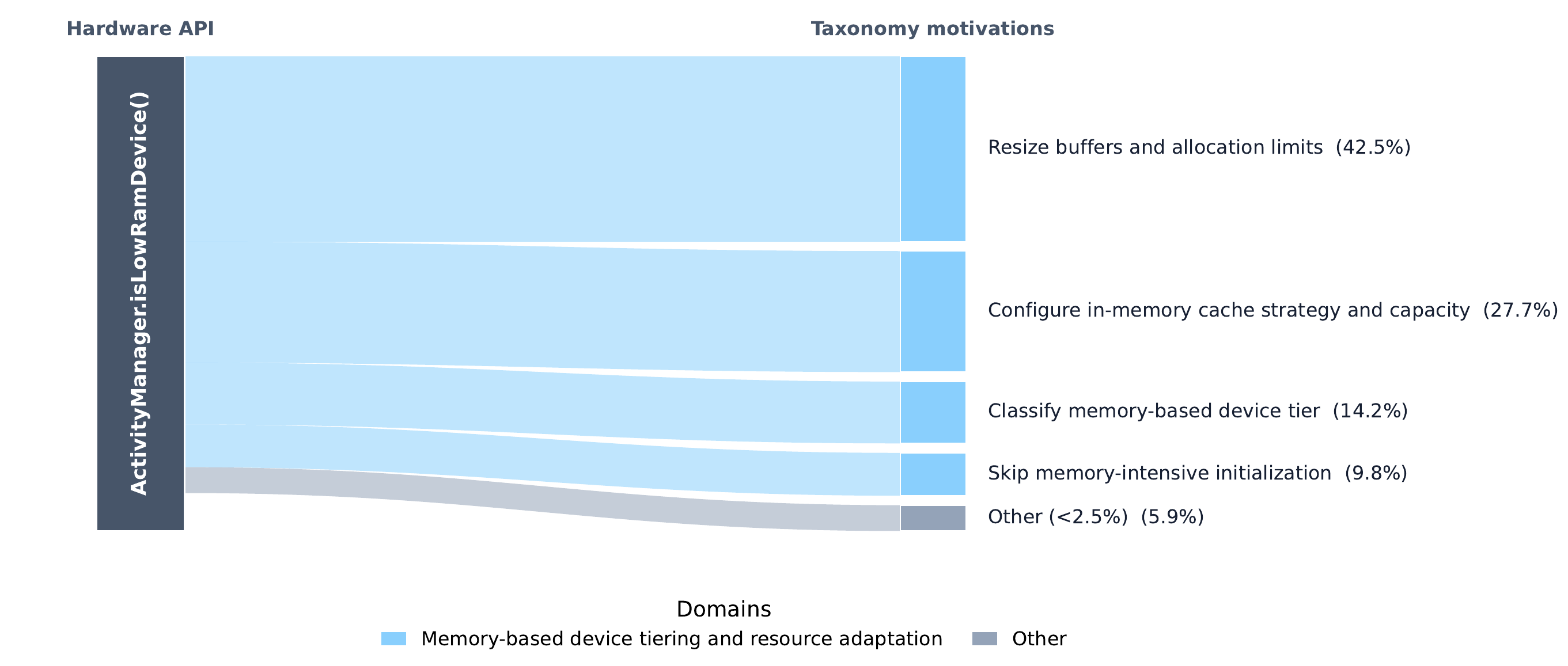}
  \caption{Sankey diagram mapping \texttt{ActivityManager.isLowRamDevice()} to taxonomy motivations.}
  \label{fig:isLowRamDeviceMotivations}
\end{figure*}

In this final RQ, we investigate how hardware-related APIs are associated with the motivations identified in RQ2, and whether an API supports a focused set of motivations or multiple distinct behaviors. We illustrate these mappings through two complementary types of Sankey diagrams. In a ``forward'' Sankey diagram, a specific hardware-related API is shown on the left and the motivations assigned to the hardware-related if statements involving that API are shown on the right. Conversely, in a ``reverse'' Sankey diagram, a specific motivation is shown on the left and the hardware-related APIs used by the corresponding if statements are shown on the right. In both cases, flow width represents the share of hardware-related if statements represented by each mapping. We present two illustrative cases: \wcircle{1} \texttt{ActivityManager.isLowRamDevice()}, a specialized API that explicitly reports whether Android classifies a device as low-RAM; and \wcircle{2} \emph{Detect emulator environment}, the most frequent individual motivation in our taxonomy (see RQ2). All other Sankey diagrams are available in our replication package but cannot be included in the paper because of space limitations. Figure~\ref{fig:isLowRamDeviceMotivations} presents the mapping from \texttt{ActivityManager.isLowRamDevice()} to the motivations assigned to the hardware-related if statements involving this API.

\begin{figure*}[ht]
  \centering
  \includegraphics[width=0.95\linewidth]{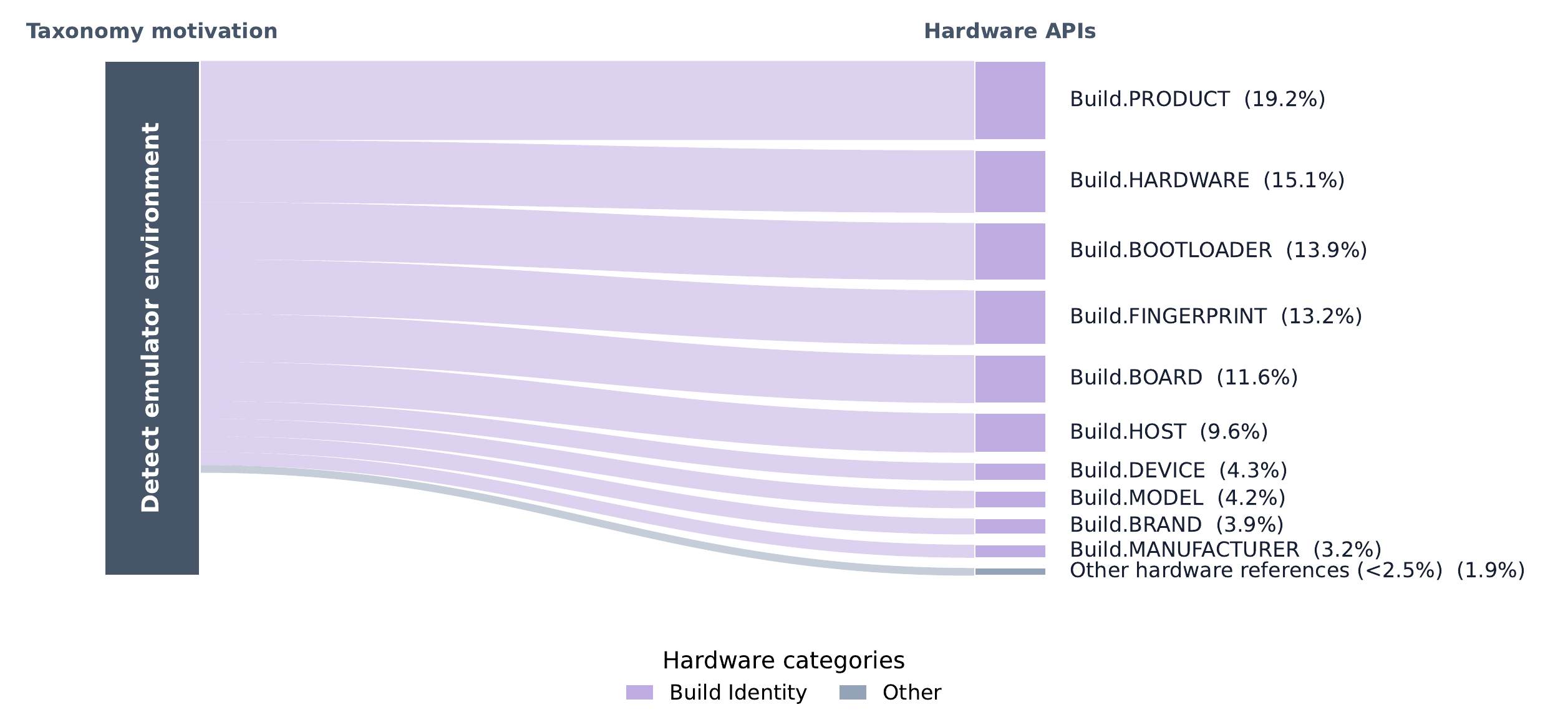}
  \caption{Sankey diagram mapping the \emph{Detect emulator environment} motivation to hardware-related APIs.}
  \label{fig:emulatorDetectionReferences}
\end{figure*}

It can be observed that \texttt{ActivityManager.isLowRamDevice()} is primarily used to resize buffers or allocation limits (\num{42.5}\%) and configure in-memory cache strategy and capacity (\num{27.7}\%), while \num{14.2}\% of its uses classify a memory-based device tier and \num{9.8}\% skip memory-intensive initialization. All of these motivations share the goal of adapting resource consumption to the device's memory profile. In practice, developers likely use this API to distinguish lower-end devices, for which memory-intensive operations may need to be constrained, from higher-end devices that can sustain larger caches, allocations, or more demanding initialization paths.  It follows that the API supports several closely related resource-management decisions rather than a single fixed adaptation.
More generally, the mappings observed in our dataset are largely intuitive: APIs that expose a specific hardware property tend to support motivations directly related to that property. For instance, memory-related APIs are mainly used for resource-management decisions, camera characteristics for camera configuration, and storage APIs for storage-aware behavior. However, this pattern is less pronounced for device-identity information. In particular, \texttt{Build.MODEL} and other \texttt{Build} properties are reused for several distinct purposes, including device-specific workarounds, vendor- or ROM-specific behavior, and emulator detection. Thus, while most hardware-related APIs have a relatively focused semantic role, build-identity APIs act as more general-purpose inputs to multiple forms of adaptation.
We next take the reverse perspective and examine the hardware references supporting the most frequent motivation, i.e., \emph{Detect emulator environment}. Figure~\ref{fig:emulatorDetectionReferences} presents a "reverse" Sankey diagram: the motivation is shown on the left, while the hardware-related APIs used by the corresponding hardware-related if statements are shown on the right.

Our analysis reveals that emulator detection in real-world apps relies on a broad set of \texttt{Build} properties. This is particularly interesting because classic security-oriented studies commonly focus on a small set of fields, especially \texttt{Build.PRODUCT}, \texttt{Build.MODEL}, and \texttt{Build.HARDWARE}~\cite{petsas2014rage,vidas2014evading}. \texttt{Build.PRODUCT} is the most frequent reference (\num{19.2}\%), typically matching emulator product images such as \texttt{sdk\_x86}, \texttt{vbox86p}, or \texttt{sdk\_gphone\_}; \texttt{Build.HARDWARE} (\num{15.1}\%) identifies virtual hardware backends through markers such as \texttt{goldfish}, \texttt{ranchu}, \texttt{vbox86}, or \texttt{qemu}. Notably, developers also use less commonly discussed properties: \texttt{Build.BOOTLOADER} (\num{13.9}\%) and \texttt{Build.BOARD} (\num{11.6}\%) to detect Nox-specific values, \texttt{Build.FINGERPRINT} (\num{13.2}\%) to match generic, unknown, or \texttt{google\_sdk} build fingerprints, and \texttt{Build.HOST} (\num{9.6}\%) to identify build-host artifacts such as \texttt{Build} or \texttt{Droid4x-BuildStation}. In contrast, the better-known \texttt{Build.MODEL} accounts for only \num{4.2}\%. Rather than relying on a single decisive property, these checks often combine complementary emulator artifacts, suggesting that analyses and tests of emulator-aware code should not restrict their attention only to the most familiar identity fields.

Outside emulator detection, most API-to-motivation mappings retain an intuitive correspondence between the behavior being implemented and the hardware property on which it depends. For example, camera adaptation is primarily supported by camera-characteristics APIs, biometric authentication by biometric APIs, and sensor-dependent features by sensor APIs. Device-identity information is a notable exception: \texttt{Build.MODEL} and related \texttt{Build} properties support several distinct forms of adaptation. For instance, we observed CameraX-based code that matches \texttt{Build.MODEL} against lists of known devices to enable a JPEG-capture downsizing quirk, thereby adapting the camera pipeline to model-specific behavior. In another case, code identifies Sony BRAVIA models through \texttt{Build.MODEL} and uses this information when selecting the display-rendering path: it retrieves the device's physical display dimensions for the matched TVs and otherwise falls back to a \num{3840} $\times$ \num{2160} resolution. These cases show that build-identity APIs can serve as general-purpose inputs to adaptation decisions beyond simple device identification.

\begin{answer}
\textbf{Answer to RQ3.}
Hardware-related APIs generally support behavior that is intuitively related to the property they expose. A notable exception is build-identity information (e.g., \texttt{Build.MODEL}, \texttt{Build.DEVICE}, etc.), which is reused across a broader range of motivations. Nevertheless, a semantically focused API does not imply a single use case. For example, \texttt{ActivityManager.isLowRamDevice()} supports multiple resource-management decisions such as cache configuration, allocation limits, initialization, and device tiering.
\end{answer}

\section{Discussion and Real-World Examples}
\label{sec:discussion}
In this section, we discuss the implications of our results and present concrete examples from the real-world apps we analyzed to illustrate how developers implement hardware-related decisions.

\noindent
\textbf{Hardware-related if statements are widespread in Android apps.}
Our experimental results show that \appsWithHardwareIfStatementsPercentage of the successfully analyzed apps contain at least one hardware-related if statement, with an average of \averageHardwareIfStatementsPerApp per app. One might expect Android's platform-level adaptation mechanisms to absorb much of device heterogeneity, rather than developers explicitly differentiating the behavior of the app in its code. Instead, explicit hardware-related if statements appear to be a recurring implementation mechanism used by developers in the majority of Android apps. This phenomenon has several implications, such as complicating dynamic app testing (e.g., automated testing on a limited set of devices may not cover all code paths), increasing the risk of compatibility issues (e.g., a new device model may trigger an untested path), and resulting in different user experiences (e.g., users with different devices may see different features or performance), among others. We will now explore some concrete examples of the patterns we observed in the real-world apps we analyzed.
To improve readability, the code listings in this section are pseudocode reconstructions of the Jimple code extracted from the analyzed APKs.

\noindent
\textbf{Low-end devices receive resource-conscious configurations.}
The observed findings show that memory-related APIs are used in many apps to select different behavior for low-memory devices, such as reducing cache sizes, lowering image quality, or disabling certain features. This indicates that developers are aware of the limitations of low-end devices and are actively trying to optimize their apps for these users. Listings~\ref{lst:lowRamCache} and~\ref{lst:lowRamScaling} show two representative cases. In a fitness app, a check of \texttt{ActivityManager.isLowRamDevice()} lowers the image-memory-cache factor to \texttt{0.15} before the cache is built. In a video-recording app, the same API selects a smaller scaling factor (\texttt{0.33} rather than \texttt{0.40}) that is subsequently used with display dimensions to compute image or buffer sizes. 

\begin{examplelisting}
{Example of low-memory scaling adaptation in a real-world video-recording app.}
{lst:lowRamScaling}
if (activityManager.isLowRamDevice())
    scaleFactor = 0.33;
else
    scaleFactor = 0.40;
bufferSize = computeSize(scaleFactor, displayWidth, displayHeight);
\end{examplelisting}

\begin{examplelisting}
{Example of image-cache configuration for a low-memory device in a real-world fitness app.}
{lst:lowRamCache}
if (activityManager.isLowRamDevice())
    imageCache.setMaxSizePercent(0.15);
else
    imageCache.setMaxSizePercent(defaultCacheFraction);
    imageCache.enableWeakCache();
\end{examplelisting}

These examples demonstrate concrete adaptations for low-end devices: constraining cache capacity and reducing the memory footprint of graphics-related allocations. This has particular implications for apps deployed in developing countries and other markets where lower-cost devices may be more prevalent, such as those in the West African Economic and Monetary Union~\cite{diallo2025security}. Indeed, in these markets, entry-level Android smartphones have been sold for around US\$40--50, making low-memory and resource-aware adaptation consequential for a substantial share of users~\cite{dalberg2017accelerating,gsma2021maisha}. The same app can therefore follow different runtime resource policies for different users, and developers targeting these markets may need to account explicitly for such device diversity to ensure that essential functionality remains usable across device classes.

\noindent
\textbf{Brand checks select vendor-specific code paths.}
Beyond simply identifying a vendor, brand and manufacturer properties allow developers to encode fine-grained workarounds for vendor-specific platform behavior, media stacks, and security implementations that cannot reliably be captured by a generic framework capability alone. Listings~\ref{lst:vendorDetachedSurface} and~\ref{lst:onePlusBiometric} show two representative cases from real-world apps. In the first, although the media stack reports support for the \texttt{detached-surface} feature, the app explicitly forces that capability to \texttt{false} on Xiaomi, OPPO, realme, and Motorola devices. In the second, a keychain component classifies a device as affected by a biometric compatibility issue only when it is a OnePlus device and its model is not part of an explicit exception list. 

\begin{examplelisting}
{Example of a vendor-specific media-capability workaround in a real-world app.}
{lst:vendorDetachedSurface}
supportsDetachedSurface = codecCapabilities.isFeatureSupported("detached-surface");
if (Build.MANUFACTURER.equals("Xiaomi")    || Build.MANUFACTURER.equals("OPPO")
    || Build.MANUFACTURER.equals("realme") || Build.MANUFACTURER.equals("motorola"))
    supportsDetachedSurface = false;
\end{examplelisting}

\begin{examplelisting}
{Example of OnePlus-specific biometric-compatibility detection in a real-world app.}
{lst:onePlusBiometric}
if (!Build.BRAND.equalsIgnoreCase("oneplus"))
    return false;
else if (ONE_PLUS_MODELS_WITHOUT_BIOMETRIC_BUG.contains(Build.MODEL))
    return false;
else
    return true;
\end{examplelisting}

These examples illustrate how vendor checks can override a framework-reported media capability or encode a precise security-related compatibility policy, rather than merely recording device identity. Such code paths may be necessary to accommodate vendor extensions and platform variations, but they also require maintenance as device families and vendor software evolve.

\noindent
\textbf{Developers may define their own device tiers to drive multiple adaptations.}
One particularly interesting pattern emerging from our taxonomy is device tiering, where developers use hardware information to assign devices to custom levels or categories rather than directly selecting a single behavior. For example, in Listing~\ref{lst:deviceTiering}, a media-player app that we analyzed uses \texttt{ActivityManager.isLowRamDevice()} to assign the explicit \texttt{DevicePerformance.HIGH\_END} or \texttt{DevicePerformance.LOW\_END} value; the high-end path also logs the chosen tier together with RAM information. 

\begin{examplelisting}
{Example of memory-based device tiering in a real-world media-player component.}
{lst:deviceTiering}
if (activityManager.isLowRamDevice())
    performance = DevicePerformance.LOW_END;
else
    performance = DevicePerformance.HIGH_END;
    log("Device performance", performance, totalRam, sdkVersion);
\end{examplelisting}

Unlike the resource-management cases discussed above, in which a memory-related check directly changes a cache, buffer, or initialization path, device tiering first translates hardware information into an app-level abstraction that can potentially guide multiple downstream decisions. It follows that memory capacity or other hardware characteristics can be translated into an internal notion of a low-, medium-, or high-end device, which may then be reused throughout the app to guide several adaptation decisions. In this sense, developers may build their own app-specific abstractions to handle device heterogeneity, rather than repeatedly checking the same hardware properties at each decision point. Our current intra-procedural analysis cannot fully capture how these classifications are propagated, combined, or reused across methods, making their construction and evolution an interesting direction for future work. A deeper interprocedural analysis could investigate whether these custom tiers are defined consistently, how widely they are reused, and how they influence the overall adaptation logic of the app.

\noindent
\textbf{Developers rely on explicit checks despite platform-level adaptation mechanisms in some cases.}
One surprising finding is that developers sometimes encode these decisions directly in the code despite platform-level mechanisms being available, such as configuration-specific resources, split APKs, or App Bundles. For instance, we observed a real-world app with an if statement condition comparing \texttt{Configuration.smallestScreenWidthDp} against the \num{600}~dp tablet threshold and explicitly classifying the device as a tablet. A layout resource placed in a directory such as \texttt{layout-sw600dp} would have allowed Android to select an alternative interface automatically for devices whose smallest available screen width is at least \num{600}~dp. This is because Android compares the current device configuration with resource-directory qualifiers at runtime and selects the best matching layout~\cite{androidAlternativeResources}. Nevertheless, the developer still prefers to handle the classification explicitly. This choice embeds the adaptation decision in imperative code rather than delegating interface selection to Android's resource system. The rationale for this explicit implementation choice cannot be determined from the code alone.

\noindent
\textbf{Hardware information supports diverse app-level decisions.}
Taken together, these use cases and our experimental findings (Section~\ref{sec:experimentalResults}) show that hardware-related if statements support a diverse set of motivations and are implemented through correspondingly diverse API choices. The real-world cases discussed above are illustrative examples and do not represent the full range of uses observed in the corpus. Other uses observed in real-world apps include building hardware-capability inventories, checking GPU and graphics support before initializing rendering components, enabling sensor-dependent interaction only when the required hardware is present, selecting architecture-compatible native code, and deferring storage-consuming operations when resources are insufficient. These examples further show that hardware information is not limited to a narrow set of device-specific workarounds, but can influence both the availability of individual features and the internal configuration of app components. Hence, hardware-related information can shape many aspects of app behavior, including feature availability, compatibility, performance, and failure handling. As a consequence, two users of the same app may follow different execution paths depending on their device's capabilities, configuration, available resources, or execution environment.
\section{Limitations and Threats to Validity}
\label{sec:limitations}
In this section, we discuss the main limitations of our study, focusing on both technical constraints and methodological challenges.

\textbf{Large-scale characterization over exhaustive program understanding.}
Our study deliberately prioritizes scalability over exhaustive program understanding. This trade-off allows us to characterize the phenomenon at the ecosystem scale and to build a broad taxonomy of motivations, while accepting that some more complex hardware-dependent behaviors may remain outside our analysis. This design choice enables the analysis of \num{50000} apps, but introduces a few main limitations:

\begin{itemize}[leftmargin=*]
    \item \textit{\wcircle{1} Intra-procedural analysis.}
    Our static analysis is intentionally intra-procedural and does not follow hardware-related information across method calls, parameters, callbacks, or other inter-procedural data flows. Similarly, when an if or else branch invokes another method, we do not recursively inspect its effects. As a consequence, we may miss checks in which hardware-related information is propagated through more complex code paths. Extending the analysis to track these flows represents an interesting direction for future work, particularly for studying patterns such as device tiering across different parts of an app.
    \item \textit{\wcircle{2} Hardware-related API inventory.}
    Our analysis is limited to the \hardwareRelatedAPIs hardware-related APIs identified through the procedure described in Section~\ref{sec:hardwareRelatedAPIs}; other relevant framework or vendor-specific APIs may therefore be omitted. An automated labeling procedure could have expanded the API list further, but it would also have introduced additional bias. Instead, as a trade-off between scalability and exhaustive coverage, we intentionally prioritize a manually validated and practically scoped list; our results should therefore be interpreted with respect to the APIs included in this list.
    \item \textit{\wcircle{3} Conditional-construct coverage.}
    Our extraction targets only Jimple \texttt{IfStmt} instructions. It does not analyze
    Jimple \texttt{TableSwitchStmt} or \texttt{LookupSwitchStmt} instructions, which
    represent Java \texttt{switch} statements. Consequently, hardware-dependent
    decisions expressed through switch cases are outside the scope of our results. We made this choice to keep the extraction lightweight at scale, since switch cases require more complex control-flow reconstruction.
    \item \textit{\wcircle{4} Static analysis only.}
    As with other static analyses, our approach may be affected by obfuscation, reflection, dynamically loaded code, native code, or constructs that cannot be reliably resolved from the APK bytecode. Moreover, we do not assess whether an identified hardware-related if statement is reachable from an app entry point, or whether either corresponding branch can be executed. The identified checks therefore represent potential program behavior rather than behavior guaranteed to occur at runtime. Dynamic analysis could provide complementary evidence, but exercising \num{50000} apps across the large number of possible hardware configurations would be impractical at the scale of our study.
\end{itemize}

\noindent
These trade-offs are consistent with our goal of providing a large-scale empirical characterization rather than a complete reconstruction of hardware-dependent behavior in each app.

\noindent
\textbf{Taxonomy Validity and subjectivity.}
There is no objective ground truth for the motivations underlying hardware-related if statements unlike "traditional" tasks evaluated against pre-existing labels. This is inherent to open coding, where categories are developed inductively from the available evidence rather than matched against a predefined set of correct labels~\cite{cruzes2011thematic,stol2016grounded}. Building and consolidating a taxonomy therefore necessarily involves human interpretation and is subject to potential subjectivity, particularly when distinguishing closely related motivations. Different researchers may reasonably interpret the same code evidence differently and derive alternative taxonomic boundaries, labels, or groupings. The same consideration applies to our hardware-related API list: there is no fully objective boundary defining which Android interfaces should be considered hardware-related, and alternative but reasonable scoping decisions could include or exclude specific APIs. We mitigate these threats by consolidating overlapping concepts before the final assignment and freezing the taxonomy during classification. To make these interpretive decisions transparent and enable scrutiny, replication, or refinement by future work, we release the taxonomy, the hardware-related API list, and the complete coding process.

\noindent
\textbf{LLM-Related Challenges.}
LLMs are used in our study only as a tool to scale the summarization of the code evidence and to support the open-coding process, without replacing the underlying Jimple evidence or the manual consolidation of the taxonomy. However, as LLMs are inherently non-deterministic, their outputs may be inconsistent or hallucinated. We mitigate this threat through evidence-bound prompts and manual taxonomy consolidation and, for the final assignment, by repeating the query 5 times and retaining the majority prediction as the final label. These measures reduce output variability and improve the robustness of the reported results, although hallucinations and prediction inconsistencies cannot be completely eliminated and may still affect the full reproducibility of the analysis.
Additionally, our experimental pipeline relies on a single open-source LLM (\texttt{gpt-oss:20b}). Although alternative open-source models and well-known large commercial models could produce somewhat different labels or individual assignments, this does not necessarily imply substantially different overall findings. As discussed in the taxonomy-validity threat, there is no single objective taxonomy for this task, since alternative but reasonable interpretations may lead to slightly different category boundaries or labels. We therefore expect that changing the model may affect some individual assignments while leaving the high-level findings of the study substantially unchanged. Accordingly, we selected a relatively small open-source model to keep the analysis scalable, reproducible, and computationally and financially manageable, rather than aiming to compare alternative LLMs. The reported results should therefore be interpreted with respect to this specific model and configuration.

\section{Related Work}
\label{sec:related}

In this section, we position our study with respect to prior work.

\noindent
\textbf{Android device-specific behavior and fragmentation.}
The most closely related work is by Dong et al., who developed a static-analysis pipeline to study device-specific behaviors in Android apps~\cite{dong2024device} and showed that such behaviors are common in apps distributed through Chinese app markets. However, unlike Dong et al., who specifically target behaviors conditioned on device identity information (e.g., the device's brand), our study considers hardware-related conditional code beyond device identity, including information not tied to a particular device family, such as memory-related APIs. On top of that, we analyze this phenomenon across \num{50000} Android apps and use the extracted branch contexts to build a taxonomy of the motivations behind hardware-related if statements. Other studies have examined Android fragmentation mainly through device-specific compatibility issues. Chen et al. analyzed \num{197} such issues and their repair practices~\cite{chen2024demystifying}, while earlier work proposed techniques such as FicFinder and PIVOT and studied evolution-induced incompatibilities and fixes~\cite{wei2016fragmentation,wei2019pivot,he2018evolution,xia2020compatibility}. Unlike these works, we study hardware-related conditional code regardless of whether it addresses a compatibility issue, capturing both device-specific workarounds and broader adaptive behaviors.

\noindent
\textbf{Device fingerprinting and emulator detection.}
A parallel line of work in security and privacy focuses on device fingerprinting, i.e., the collection of device properties to distinguish or recognize a device, as well as on signals for detecting emulation environments. AndroPRINT automatically identifies Android API sources that can contribute to a device fingerprint~\cite{palfinger2020androprint}; BareDroid enables large-scale analysis on physical devices because malware can recognize emulated environments~\cite{mutti2015baredroid}; and Kondracki et al. show that configuration, user-data, and sensor artifacts available through Android APIs can distinguish real devices from sandboxes~\cite{kondracki2022droid}. Recent large-scale work also maps fingerprinting-like behavior in third-party mobile SDKs~\cite{specter2025fingerprinting}. These works are closely related to ours because they also rely on similar Android-exposed device information, such as build, configuration, and sensor information. However, they study these signals mainly for fingerprinting or emulator detection, whereas our work examines them more broadly as inputs to hardware-related if statements, with emulator detection representing only one motivation in our taxonomy.

\section{Conclusion}
\label{sec:conclusion}
Android apps run across a highly heterogeneous device ecosystem, yet how developers use hardware-related information to adapt their behavior remains poorly understood. In this paper, we conducted a large-scale study of hardware-related conditional code in \num{50000} Android apps, finding that \appsWithHardwareIfStatementsPercentage of successfully analyzed apps contain at least one such statement, with an average of \averageHardwareIfStatementsPerApp per app. We then used LLM-assisted open coding to build a taxonomy comprising \taxonomyMotivations motivations organized into \taxonomyDomains domains, spanning feature gating, UI adaptation, emulator detection, compatibility handling, resource management, device tiering, and other forms of hardware-related behavior. Our findings provide a first systematic view of how hardware-related information is used by developers and open several directions for future work, including studying how hardware-related information propagates beyond local hardware-related if statements through interprocedural analysis and how higher-level strategies, such as device tiering, are encoded and reused throughout app code.

\section*{Data Availability}
\label{sec:dataAvailability}
We release our taxonomy and all supporting research artifacts to the community to facilitate reproducibility and further research:
\begin{center}
   \url{https://anonymous.4open.science/r/AndroidHardwareAPIs-777}
\end{center}

\bibliographystyle{ACM-Reference-Format}
\bibliography{references}

\end{document}